\documentclass[preprint]{aastex}

\shorttitle{}

\usepackage{lineno}
\usepackage{amsmath}

\newcommand{\tightlist}{
 \begin{list}{$\bullet$}
  { \setlength{\itemsep}{0pt}
     \setlength{\parsep}{3pt}
     \setlength{\topsep}{3pt}
     \setlength{\partopsep}{0pt}
     \setlength{\leftmargin}{1.5em}
     \setlength{\labelwidth}{1em}
     \setlength{\labelsep}{0.5em} } }

\newcommand{\listend}{
  \end{list}  }

\begin{document}

\title{Search for muon neutrinos from Gamma-Ray Bursts with the
IceCube neutrino telescope}

\author{
IceCube Collaboration:
R.~Abbasi\altaffilmark{1},
Y.~Abdou\altaffilmark{2},
T.~Abu-Zayyad\altaffilmark{3},
J.~Adams\altaffilmark{4},
J.~A.~Aguilar\altaffilmark{1},
M.~Ahlers\altaffilmark{5},
K.~Andeen\altaffilmark{1},
J.~Auffenberg\altaffilmark{6},
X.~Bai\altaffilmark{7},
M.~Baker\altaffilmark{1}
% ...
}
\altaffiltext{1}{Dept.~of Physics, University of Wisconsin, Madison, WI 53706, USA}
\altaffiltext{2}{Dept.~of Subatomic and Radiation Physics, University of Gent, B-9000 Gent, Belgium}
\altaffiltext{3}{Dept.~of Physics, University of Wisconsin, River Falls, WI 54022, USA}
\altaffiltext{4}{Dept.~of Physics and Astronomy, University of Canterbury, Private Bag 4800, Christchurch, New Zealand}
\altaffiltext{5}{Dept.~of Physics, University of Oxford, 1 Keble Road, Oxford OX1 3NP, UK}
\altaffiltext{6}{Dept.~of Physics, University of Wuppertal, D-42119 Wuppertal, Germany}
\altaffiltext{7}{Bartol Research Institute and Department of Physics and Astronomy, University of Delaware, Newark, DE 19716, USA}
% ...

\begin{abstract}
We present the results of searches for high-energy muon neutrinos from
41 gamma-ray bursts (GRBs) in the northern sky with the IceCube
detector in its 22-string configuration active in 2007/2008. The
searches cover both the prompt and a possible precursor emission as
well as a model-independent, wide time window of $-1\,$h to $+3$\,h
around each GRB. In contrast to previous searches with a large GRB
population, we do not utilize a standard Waxman--Bahcall GRB flux for
the prompt emission but calculate individual neutrino spectra for all
41 GRBs from the burst parameters measured by satellites. For all
three time windows the best estimate for the number of signal events
is zero. Therefore, we place 90\% CL upper limits on the fluence from
the prompt phase of $3.7 \times
10^{-3}\,\mathrm{erg}\,\mathrm{cm}^{-2}$ (72\,TeV -- 6.5\,PeV) and on
the fluence from the precursor phase of $2.3 \times
10^{-3}\,\mathrm{erg}\,\mathrm{cm}^{-2}$ (2.2\,TeV -- 55\,TeV), where
the quoted energy ranges contain 90\% of the expected signal events in
the detector. The 90\% CL upper limit for the wide time window is $2.7
\times 10^{-3}\,\mathrm{erg}\,\mathrm{cm}^{-2}$ (3\,TeV -- 2.8\,PeV)
assuming an $E^{-2}$ flux.
\end{abstract}


\section{Introduction}
Gamma-ray bursts (GRBs) are among the most violent events in the
universe and among the few plausible candidates for sources of the
ultra-high energy cosmic rays. So-called long-duration GRBs ($\gtrsim
2$\,s) are thought to originate from the collapse of a massive star
into a black hole \citep{apj:405:273}, whereas short-duration GRBs
($\lesssim 2$\,s) are believed to be the result of the merger of two
compact objects (e.g., neutron stars) into a black hole
\citep{nat:340:126}. Though quite different in nature both scenarios are
consistent with the currently leading model for GRBs, the fireball
model \citep{apj:405:278}, with the energy source (central engine)
being the rapid accretion of a large mass onto the newly formed black
hole. In this model, a highly relativistic outflow (fireball)
dissipates its energy via synchrotron or inverse Compton radiation of
electrons accelerated in internal shock fronts
\citep{apj:395:l83,apj:430:l93,apj:485:270}. This radiation in the
keV--MeV range is observed as the $\gamma$-ray signal. In case of
long GRBs the energy in gamma rays is typically of ${\cal
O}(10^{51}$--$10^{54}$\,erg$\,\times\,\Omega / 4\pi)$ where $\Omega$
is the opening angle for the $\gamma$-ray emission. Short GRBs are
observed to release about a factor 100 less energy.

In addition to electrons, protons are thought to be accelerated via
the Fermi mechanism, resulting in an $E^{-2}$ power law spectrum with
energies up to $10^{20}$\,eV \citep{prl:75:386,apj:453:883}. The
normalization of the proton spectrum is usually given in relation to
the energy in electrons. The latter is linked to the energy in
$\gamma$-ray photons through the synchrotron and inverse Compton
energy-loss mechanisms. Protons of ${\cal O}(10^{15}\,\mathrm{eV})$
interact with the keV--MeV photons forming a $\Delta^+$ resonance
which decays into pions \citep{prl:78:2292}. In the decay of the
charged pions, neutrinos of energy ${\cal O}(10^{14}\,\mathrm{eV})$
are produced with the approximate ratios
($\nu_e$:$\nu_\mu$:$\nu_\tau$) = (1:2:0)
\footnote{Here and throughout the rest of the paper $\nu$ denotes both
neutrinos and antineutrinos.}, changing to about (1:1:1) at Earth due
to oscillations \citep{app:1995:3:267,mpl:a21:1049}.  First calculations of
this prompt neutrino flux
\citep{prl:78:2292,apj:521:928} used average GRB parameters and the
GRB rate measured by BATSE to determine an all-sky neutrino flux from
the GRB population. The AMANDA-II neutrino telescope
\citep{apj:664:397,apj:674:357} performed searches for this so-called
Waxman--Bahcall GRB flux or similar GRB fluxes
\citep{apj:664:397,apj:674:357} with negative results.

In a similar way, so-called precursor neutrinos can be generated when
the expanding fireball is still inside the progenitor star
\citep{pr:d68:083001}. In this case, the accelerated protons interact
with matter of the progenitor star or synchrotron photons. However,
due to the large optical depth the synchrotron photons cannot escape
the fireball and, hence, no $\gamma$-ray signal is observed. The time
delay between the start of this neutrino emission and the prompt
$\gamma$-ray signal is expected to be about 100\,s.

Observations of the early and late afterglow phases reveal that a
large fraction of GRBs show $X$-ray flares superposed on the decaying
light curve. Sometimes these flares are interpreted as a restart of
the central engine that already generated the prompt emission
\citep{ptrsl:a365:1213}. If this is true, neutrino production with a
similar spectrum as the prompt emission can be expected in the
afterglow phase up to $10^4\,$s after the $\gamma$-ray signal
\citep{prl:97:051101}. Furthermore, production of neutrinos with
energies around $10^{18}\,$GeV is expected when the shock fronts
collide with the interstellar medium or the progenitor wind \citep{apj:541:707}.

In our analysis we search for muon neutrinos from GRBs recorded by
satellites between 2007 June 1 and 2008 April 4 in all three
phases. For the prompt phase we utilize both an unbinned likelihood
and a binned method. We find that the unbinned likelihood method has a
significantly better discovery potential and is therefore used to
obtain the limits presented in this paper.  For searches in other
emission phases we perform only an unbinned search.  The paper is
structured as follows: In Section~\ref{sec:speccalc} we define the
neutrino spectra used for the different phases, followed by a
description of the IceCube detector in
Section~\ref{sec:detector}. Afterwards, in Sections~\ref{sec:data} and
\ref{sec:simulation} the data sets and simulations are discussed,
respectively. In Section~\ref{sec:methodology} the unbinned likelihood
and binned methods are described and their performance is
compared. Section~\ref{sec:results} then presents the results followed
by a discussion of systematic uncertainties. Finally,
Section~\ref{sec:otherResults} sets the results into context with
other observations.

\section{GRB neutrino-spectra and time windows}\label{sec:speccalc} 

The searches in this paper rely on the directional, temporal and
spectral information obtained from satellite-based $\gamma$-ray
observations which are distributed via the Gamma-ray burst Coordinate
Network (GCN, \citet{web:gcn:homepage}). Primarily, this information
comes from \emph{Swift} \citep{ssr:120:165} (also $X$-ray and UV
observations), but also from Konus-Wind \citep{wind:homepage},
SuperAGILE \citep{Tavani:2008zz}, Integral \citep{Mereghetti:2003fr},
and other satellites of the Third Interplanetary Network
\citep{ipn:homepage}.  In our analyses we only consider bursts which
occurred at a declination above $-5^\circ$.  The southern sky is
dominated by downgoing muons created by cosmic ray interactions with
the atmosphere. By restricting our searches to the northern sky, the
background from downward going muons is drastically reduced. The
resolution of the GRB position from the satellites is better than
$0.1^\circ$, well below the resolution of the IceCube detector. It is
therefore neglected in these analyses.

\subsection{Prompt emission}\label{sec:nuFluxesPrompt}
We calculate the expected prompt neutrino spectrum in the internal
shock scenario of the fireball model following \citet{app:20:429}
which is based on \citet{prl:78:2292}. For reference we list all
formulae used in our calculations in Appendix \ref{app:nu_spec}
together with a definition of the various parameters.  Prompt neutrino
emission from GRBs is the result of meson production in collisions of
accelerated protons and the observed $\gamma$-rays in the keV--MeV
range. It is therefore expected to occur during the same time frame as
the $\gamma$-ray emission and to track the photon energy spectrum. This
is reflected in the similar functional form of $F_\gamma$ (Equation
(\ref{eq:photonspec})) and $F_\nu$ (Equation
(\ref{eq:nuspec})). Due to the $\Delta^+$ resonance condition, the
neutrino energy is predicted to be inversely proportional to the
photon energy, which is illustrated in the definition of $\alpha_\nu$
and $\beta_\nu$ in Equation (\ref{eq:nuIndices}). The further break in
the neutrino spectrum above an energy $\epsilon_2$ is due to
synchrotron cooling of high energy pions and muons before producing
neutrinos. The expected energy fluence in neutrinos is directly
proportional to the measured energy fluence in photons (Equation
(\ref{eq:nuIntegral})). Here, some of the measured photon indices lead
to diverging $\gamma$-ray spectral integrals if integrated from zero to
infinity in energy. As the photon spectrum will not follow a broken
power law spectrum to arbitrarily high or low energies, we limit the
integration range for all GRBs from 1\,keV to 10\,MeV for which broken
power law spectra have been observed by $\gamma$-ray satellites.

\begin{deluxetable}{lrrrrrlp{0mm}rlllp{0mm}rrrrrr}
\tabletypesize{\tiny}
\rotate
\tablecaption{Burst Parameters, $\gamma$-Ray and Neutrino Spectra of All 41
GRBs (for Definitions of Parameters See Appendix \ref{app:nu_spec})
\label{tab:grbSpectra}}
\tablewidth{0pt}
\tablehead{
& \multicolumn{6}{c}{General Burst Parameters} 
&& \multicolumn{4}{c}{$\gamma$-Ray Spectrum}
&& \multicolumn{6}{c}{$\nu$ Spectrum}\\
& \colhead{$T_0$}
& \colhead{R.A.}
& \colhead{Decl.}
& \colhead{$T_1-T_0$}
& \colhead{$T_2-T_0$}
& \colhead{$z$}
&& \colhead{$f_\gamma$}
& \colhead{$\epsilon_\gamma$}
& \colhead{$\alpha_\gamma$}
& \colhead{$\beta_\gamma$}
&& \colhead{$f_\nu$}
& \colhead{$\epsilon_{\nu,1}$}
& \colhead{$\epsilon_{\nu,2}$}
& \colhead{$\alpha_\nu$}
& \colhead{$\beta_\nu$}
& \colhead{$\gamma_\nu$}
}
\startdata
GRB 070610&20:52:26&298.8&26.2&$-0.8$&$+4.4$&2.00\tablenotemark{*}&&2.3e$+$00&0.20\tablenotemark{*}&1.76&2.76\tablenotemark{*}&&2.2e$-$17&0.35&8.54&0.24&1.24&3.24\\ 
GRB 070612A&02:38:45&121.4&37.3&$-4.7$&$+418.0$&2.00\tablenotemark{*}&&1.1e$+$02&0.20\tablenotemark{*}&1.69&2.69\tablenotemark{*}&&1.2e$-$15&0.35&8.54&0.31&1.31&3.31\\ 
GRB 070616&16:29:33&32.2&56.9&$-2.6$&$+602.2$&2.00\tablenotemark{*}&&2.2e$+$02&0.20\tablenotemark{*}&1.61&2.61\tablenotemark{*}&&2.8e$-$15&0.35&8.54&0.39&1.39&3.39\\ 
GRB 070704&20:05:57&354.7&66.3&$-57.3$&$+400.8$&2.00\tablenotemark{*}&&5.4e$+$01&0.20\tablenotemark{*}&1.79&2.79\tablenotemark{*}&&4.9e$-$16&0.35&8.54&0.21&1.21&3.21\\ 
GRB 070714B&04:59:29&57.8&28.3&$-0.8$&$+65.6$&0.92&&1.1e$+$01&0.20\tablenotemark{*}&1.36&2.36\tablenotemark{*}&&5.1e$-$17&0.85&13.34&0.64&1.64&3.64\\ 
GRB 070724B&23:25:09&17.6&57.7&$-2.0$&$+120.0$&2.00\tablenotemark{*}&&1.1e$+$03&0.08&1.15&3.15&&1.8e$-$15&0.85&8.54&-0.15&1.85&3.85\\ 
GRB 070808&18:28:00&6.8&1.2&$-0.7$&$+41.4$&2.00\tablenotemark{*}&&1.6e$+$01&0.20\tablenotemark{*}&1.47&2.47\tablenotemark{*}&&2.8e$-$16&0.35&8.54&0.53&1.53&3.53\\ 
GRB 070810B&15:19:17&9.0&8.8&$+0.0$&$+0.1$&2.00\tablenotemark{*}&&1.4e$-$01&0.20\tablenotemark{*}&1.44&2.44\tablenotemark{*}&&2.6e$-$18&0.35&8.54&0.56&1.56&3.56\\ 
GRB 070917&07:33:57&293.9&2.4&$-0.1$&$+11.4$&2.00\tablenotemark{*}&&3.7e$+$01&0.21&1.36&3.36&&6.0e$-$16&0.33&8.54&-0.36&1.64&3.64\\ 
GRB 070920A&04:00:13&101.0&72.3&$+15.1$&$+75.0$&2.00\tablenotemark{*}&&5.3e$+$00&0.20\tablenotemark{*}&1.69&2.69\tablenotemark{*}&&5.7e$-$17&0.35&8.54&0.31&1.31&3.31\\ 
GRB 071003&07:40:55&301.9&10.9&$-7.6$&$+167.4$&2.00\tablenotemark{*}&&3.0e$+$01&0.80&0.97&2.97&&3.8e$-$14&0.09&8.54&0.03&2.03&4.03\\ 
GRB 071008&21:55:56&151.6&44.3&$-11.0$&$+14.0$&2.00\tablenotemark{*}&&1.2e$+$00&0.20\tablenotemark{*}&2.23&3.23\tablenotemark{*}&&6.4e$-$18&0.35&8.54&-0.23&0.77&2.77\\ 
GRB 071010B&20:45:47&150.5&45.7&$-35.7$&$+24.1$&0.95&&2.1e$+$03&0.03&1.25&2.65&&5.0e$-$17&5.71&13.16&0.35&1.75&3.75\\ 
GRB 071010C&22:20:22&338.1&66.2&$-2.0$&$+20.0$&2.00\tablenotemark{*}&&3.2e$+$01\tablenotemark{*}&0.20\tablenotemark{*}&1.00\tablenotemark{*}&2.00\tablenotemark{*}&&1.8e$-$15&0.35&8.54&1.00&2.00&4.00\\ 
GRB 071011&12:40:13&8.4&61.1&$-9.5$&$+63.8$&2.00\tablenotemark{*}&&3.2e$+$01&0.20\tablenotemark{*}&1.41&2.41\tablenotemark{*}&&6.3e$-$16&0.35&8.54&0.59&1.59&3.59\\ 
GRB 071013&12:09:19&279.5&33.9&$-5.9$&$+23.4$&2.00\tablenotemark{*}&&3.7e$+$00&0.20\tablenotemark{*}&1.60&2.60\tablenotemark{*}&&4.8e$-$17&0.35&8.54&0.40&1.40&3.40\\ 
GRB 071018&08:37:41&164.7&53.8&$-50.0$&$+417.7$&2.00\tablenotemark{*}&&1.1e$+$01&0.20\tablenotemark{*}&1.63&2.63\tablenotemark{*}&&1.4e$-$16&0.35&8.54&0.37&1.37&3.37\\ 
GRB 071020&07:02:27&119.7&32.9&$-3.0$&$+7.4$&2.15&&2.6e$+$01&0.32&0.65&2.65&&5.3e$-$15&0.20&8.13&0.35&2.35&4.35\\ 
GRB 071021&09:41:33&340.6&23.7&$-31.4$&$+252.2$&2.00\tablenotemark{*}&&1.3e$+$01&0.20\tablenotemark{*}&1.70&2.70\tablenotemark{*}&&1.4e$-$16&0.35&8.54&0.30&1.30&3.30\\ 
GRB 071025&04:08:54&355.1&31.8&$+38.5$&$+193.8$&2.00\tablenotemark{*}&&5.9e$+$01&0.20\tablenotemark{*}&1.79&2.79\tablenotemark{*}&&5.4e$-$16&0.35&8.54&0.21&1.21&3.21\\ 
GRB 071028A&17:41:01&119.8&21.5&$+0.0$&$+48.9$&2.00\tablenotemark{*}&&2.4e$+$00&0.20\tablenotemark{*}&1.87&2.87\tablenotemark{*}&&2.0e$-$17&0.35&8.54&0.13&1.13&3.13\\ 
GRB 071101&17:53:46&48.2&62.5&$-1.9$&$+10.0$&2.00\tablenotemark{*}&&3.7e$-$01&0.20\tablenotemark{*}&2.25&3.25\tablenotemark{*}&&2.1e$-$18&0.35&8.54&-0.25&0.75&2.75\\ 
GRB 071104&11:41:23&295.6&14.6&$-5.0$&$+17.0$&2.00\tablenotemark{*}&&3.2e$+$01\tablenotemark{*}&0.20\tablenotemark{*}&1.00\tablenotemark{*}&2.00\tablenotemark{*}&&1.8e$-$15&0.35&8.54&1.00&2.00&4.00\\ 
GRB 071109&20:36:05&289.9&2.0&$-5.0$&$+35.0$&2.00\tablenotemark{*}&&3.2e$+$01\tablenotemark{*}&0.20\tablenotemark{*}&1.00\tablenotemark{*}&2.00\tablenotemark{*}&&1.8e$-$15&0.35&8.54&1.00&2.00&4.00\\ 
GRB 071112C&18:32:57&39.2&28.4&$-5.0$&$+30.0$&0.82&&6.3e$+$01&0.20\tablenotemark{*}&1.09&2.09\tablenotemark{*}&&4.3e$-$16&0.95&14.07&0.91&1.91&3.91\\ 
GRB 071118&08:57:17&299.7&70.1&$-25.0$&$+110.0$&2.00\tablenotemark{*}&&5.6e$+$00&0.20\tablenotemark{*}&1.63&2.63\tablenotemark{*}&&6.8e$-$17&0.35&8.54&0.37&1.37&3.37\\ 
GRB 071122&01:23:25&276.6&47.1&$-29.4$&$+47.3$&1.14&&5.4e$+$00&0.20\tablenotemark{*}&1.77&2.77\tablenotemark{*}&&1.7e$-$17&0.69&11.97&0.23&1.23&3.23\\ 
GRB 071125&13:56:42&251.2&4.5&$-0.5$&$+8.5$&2.00\tablenotemark{*}&&3.4e$+$02&0.30&0.62&3.10&&3.8e$-$14&0.24&8.54&-0.10&2.38&4.38\\ 
GRB 080121&21:29:55&137.2&41.8&$-0.4$&$+0.4$&2.00\tablenotemark{*}&&7.9e$-$02&0.20\tablenotemark{*}&2.60&3.60\tablenotemark{*}&&3.6e$-$19&0.35&8.54&-0.60&0.40&2.40\\ 
GRB 080205&07:55:51&98.3&62.8&$-10.1$&$+105.3$&2.00\tablenotemark{*}&&1.3e$+$01&0.20\tablenotemark{*}&2.08&3.08\tablenotemark{*}&&8.0e$-$17&0.35&8.54&-0.08&0.92&2.92\\ 
GRB 080211&07:23:39&44.0&60.0&$-10.0$&$+50.0$&2.00\tablenotemark{*}&&1.3e$+$02&0.35&0.61&2.62&&3.1e$-$14&0.20&8.54&0.38&2.39&4.39\\ 
GRB 080218A&20:08:43&355.9&12.2&$-12.8$&$+18.6$&2.00\tablenotemark{*}&&2.6e$+$00&0.20\tablenotemark{*}&2.34&3.34\tablenotemark{*}&&1.3e$-$17&0.35&8.54&-0.34&0.66&2.66\\ 
GRB 080307&11:23:30&136.6&35.1&$+1.7$&$+146.1$&2.00\tablenotemark{*}&&8.0e$+$00&0.20\tablenotemark{*}&1.78&2.78\tablenotemark{*}&&7.4e$-$17&0.35&8.54&0.22&1.22&3.22\\ 
GRB 080310&08:37:58&220.1&-0.2&$-71.8$&$+318.7$&2.43&&9.6e$+$00&0.20\tablenotemark{*}&2.32&3.32\tablenotemark{*}&&7.2e$-$17&0.27&7.47&-0.32&0.68&2.68\\ 
GRB 080315&02:25:01&155.1&41.7&$-5.0$&$+65.0$&2.00\tablenotemark{*}&&4.3e$-$01&0.20\tablenotemark{*}&2.51&3.51\tablenotemark{*}&&2.0e$-$18&0.35&8.54&-0.51&0.49&2.49\\ 
GRB 080319C&12:25:56&259.0&55.4&$-0.3$&$+51.2$&1.95&&1.5e$+$02&0.11&1.01&1.87&&1.8e$-$15&0.68&8.68&1.13&1.99&3.99\\ 
GRB 080319D&17:05:09&99.5&23.9&$+0.0$&$+50.0$&2.00\tablenotemark{*}&&2.4e$+$00&0.20\tablenotemark{*}&1.92&2.92\tablenotemark{*}&&1.8e$-$17&0.35&8.54&0.08&1.08&3.08\\ 
GRB 080320&04:37:38&177.7&57.2&$-60.0$&$+40.0$&2.00\tablenotemark{*}&&2.8e$+$00&0.20\tablenotemark{*}&1.70&2.70\tablenotemark{*}&&2.9e$-$17&0.35&8.54&0.30&1.30&3.30\\ 
GRB 080325&04:09:17&277.9&36.5&$-29.3$&$+170.5$&2.00\tablenotemark{*}&&5.2e$+$01&0.20\tablenotemark{*}&1.68&2.68\tablenotemark{*}&&5.7e$-$16&0.35&8.54&0.32&1.32&3.32\\ 
GRB 080328&08:03:04&80.5&47.5&$-2.2$&$+117.5$&2.00\tablenotemark{*}&&1.0e$+$02&0.28&1.13&3.13&&5.4e$-$15&0.25&8.54&-0.13&1.87&3.87\\ 
GRB 080330&03:41:16&169.3&30.6&$-0.5$&$+71.9$&1.51&&1.0e$+$00&0.20\tablenotemark{*}&2.53&3.53\tablenotemark{*}&&3.1e$-$18&0.50&10.20&-0.53&0.47&2.47\\ 
\enddata
\tablecomments{Columns: $T_0$ -- trigger time of satellite [UT], RA -- right ascention of
GRB [$^\circ$], Dec -- declination of GRB [$^\circ$], $T_1-T_0$ --
start of prompt window [s], $T_2-T_0$ -- end of prompt window [s],
$f_\gamma$ [MeV$^{-1}$\,cm$^{-2}$], $\epsilon_\gamma$ [MeV], $f_\nu$
[GeV$^{-1}$\,cm$^{-2}$], $\epsilon_{\nu,1}$ [PeV], $\epsilon_{\nu,2}$
[PeV]. The parameters $f_\gamma$ and $f_\nu$ are the fluxes at
$\epsilon_\gamma$ and $\epsilon_{\nu,1}$ of the gamma-ray and neutrino
spectrum, respectively (see also Appendix \ref{app:nu_spec}).}
\tablenotetext{*}{Parameter has not been measured. Instead, an average value is
used (see Table \ref{tab:avGrbVal}).}  \end{deluxetable}

\begin{deluxetable}{lp{2cm}c}
\tablewidth{0pt}
\tablecaption{Average Values of GRB Parameters Taken from \citet{pr:458:173}
\label{tab:avGrbVal}}
\tablehead{
\colhead{Parameter}
&& \colhead{Average Value} }
\startdata
$f_\gamma$              && $1.3\,\mathrm{MeV}^{-1}\,\mathrm{cm}^{-2}$ \tablenotemark{a}\\
$z$                     && 2\\
$\epsilon_\gamma$       && 0.2\,MeV\\
$\alpha_\gamma$         && 1\\
$\beta_\gamma$          && $\alpha_\gamma + 1$\\
$L_\gamma^\mathrm{iso}$\tablenotemark{*} && $10^{51}\,\mathrm{erg}\,\mathrm{cm}^{-2}$\\
$\Gamma_\mathrm{jet}$\tablenotemark{*}   && 300 \\
$t_\mathrm{var}$\tablenotemark{*}        && 0.01\,s\\
$\epsilon_e$\tablenotemark{*}            && 0.1\\
$\epsilon_B$\tablenotemark{*}            && 0.1\\
$f_e$\tablenotemark{*}                   && 0.1\\
\enddata
\tablecomments{For parameter definitions see Appendix.}
\tablenotetext{a}{Corresponds to ${\cal F_\gamma} =
10^{-5}\,\mathrm{erg}\,\mathrm{cm}^{-2}$ between 10\,keV
and 10\,MeV.} 
\tablenotetext{*}{Not measured for any GRB.}
\end{deluxetable}

\begin{figure}[t]
\epsscale{.7}
\plotone{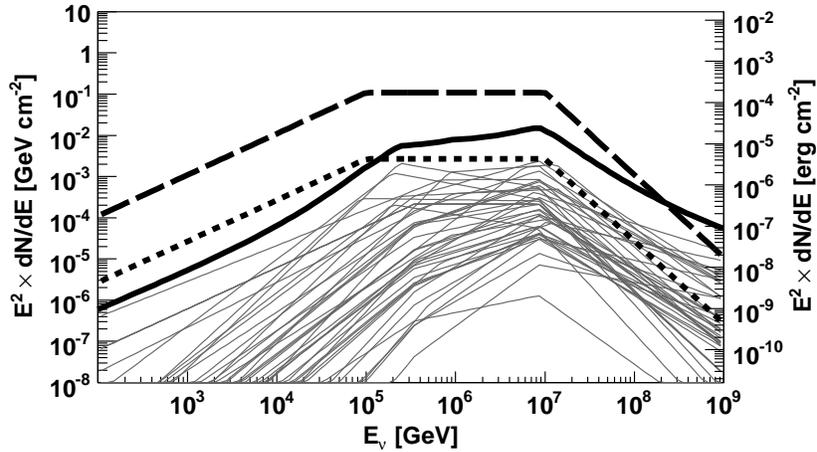}
\caption{Calculated neutrino spectra for all 41 GRBs (thin solid
  lines) compared to the standard Waxman--Bahcall
  spectrum for a single burst (thick dotted line). Also shown are the
  sum of all 41 individual spectra (thick solid line) and the sum of
  41 Waxman--Bahcall-like spectra (thick dashed line).\label{fig:nuFluxes}}
\end{figure}

The time window of the prompt emission is determined from the
information published in the GCN circulars and reports and checked
against the measured $\gamma$-ray emission curves (available from
\citet{web:gcn:homepage}). In case of early or late emission outside
the window specified by the satellite experiments the window is
extended accordingly. The exact window definitions for the bursts are
listed in Table\,\ref{tab:grbSpectra}. In previous publications on
stacked searches for neutrinos from GRBs
\citep{apj:664:397,apj:674:357} it was assumed that the sum of all GRB
spectra follows a Waxman--Bahcall GRB spectrum
\citep{prl:78:2292}. However, the burst parameters can vary
significantly from burst to burst \citep{app:20:429,app:25:118} and
the GRB population used here (mostly observed by the \emph{Swift} satellite)
is different from the BATSE population on which the calculations from
Waxman and Bahcall were based. Therefore, we use the measured
parameters to calculate the neutrino spectra for each GRB individually
as has already been done in the case of single, bright bursts
\citep{proc:icrc05:stamatikos:1,apj:subm:1}. The parameters (see
appendix \ref{app:nu_spec}) for each
GRB are listed in Table\,\ref{tab:grbSpectra}. In case a parameter has
not been measured for a particular burst the average value listed in
Table~\ref{tab:avGrbVal} is used. The table also contains assumed
values of parameters that have not been measured for any of the GRBs.
The resulting neutrino spectra for all 41 GRBs are shown in
Figure~\ref{fig:nuFluxes} together with the standard Waxman--Bahcall
spectrum\footnote{The fluence of a standard Waxman--Bahcall burst is
calculated from the flux quoted in \citet{npbps:118:353} (Equation (17),
given as a ``diffuse'' all-sky flux in
$\mathrm{GeV}^{-1}\,\mathrm{sr}^{-1}\,\mathrm{s}^{-1}$) by multiplying
it with $4\pi\,\mathrm{sr} \times 1\,\mathrm{yr}$ and dividing it by
the assumed number of bursts per year (667). The resulting fluence is
divided by two to account for neutrino oscillations (full mixing
assumed).\label{foot:wbburst}}. Clear differences in the shapes of the
two summed spectra are observed together with an ${\cal O}(10)$ times
lower overall fluence for the individual spectra. The latter is caused
by the much higher sensitivity of \emph{Swift} compared to BATSE. The
observed differences stress the importance of using individual
fluences in analyses.

\subsection{Precursor emission} 
In the case of precursor emission, the neutrino-producing interactions
occur while the fireball is still opaque to electromagnetic
emission. Therefore, no analogous photons are observed and modeling of
the emission on a per burst basis is not possible. We therefore use
the fluence derived by \citet{pr:d68:083001} for H-progenitor stars
and assume that all 41 GRBs have such a precursor phase with the same
fluence. The spectrum is significantly softer than that of the prompt
emission. Below 10\,TeV, it follows an $E^{-2}$ power law spectrum and
has a sharp drop around 60\,TeV (see Figure~\ref{fig:limits}). Neutrinos
below 60\,TeV are mainly produced in interactions of accelerated
protons with cold stellar protons whereas those above 60\,TeV originate
from proton interactions with photons in the jet. The time window is
taken as the 100\,s immediately preceding the prompt time window.  The
window is chosen to be large enough to encompass the predicted
emission and a potential delay between the phases.

\subsection{Wide window emission} 
While specific predictions have been made for neutrino emission both
before, during, and after the observed $\gamma$-ray emission of GRBs,
there are many unknown quantities that factor into the calculation of
fluence. It is therefore important to search for generic emission of
high energy neutrinos in a reasonable time window surrounding the
observed bursts. For the first time, we perform such a search in a
wide time window ($-1$\,h to $+3$\,h) around each burst. The size of
the window is motivated by possible precursor and afterglow emission,
and limited by the requirement to keep backgrounds low. Rather than
attempt to model the emission, we assume a generic $E^{-2}$ energy
spectrum. Such a spectrum is in agreement with the assumed parent
cosmic-ray spectrum and is distinguishable from the atmospheric
neutrino background (see Section~\ref{sec:simulation}).

\section{Detector and data acquisition}\label{sec:detector}

IceCube \citep{app:26:155}, the successor of the AMANDA experiment and
the first next-generation neutrino telescope, is currently being
installed in the deep ice at the geographic South Pole. Its final
configuration will instrument a volume of about $1\,\mathrm{km}^3$ of
clear ice in depths between 1450\,m and 2450\,m. Neutrinos are
reconstructed by detecting the Cherenkov light from charged secondary
particles, which are produced in interactions of the neutrinos with
the nuclei in the ice or the bedrock below. The optical sensors, known as
Digital Optical Modules (DOMs), consist of a 25\,cm Hamamatsu
photomultiplier tube (PMT) housed in a pressure-resistant glass sphere
and associated electronics \citep{nim:a601:294}. They are mounted on
vertical strings where each string carries 60 DOMs. The final detector
will contain 86 such strings spaced horizontally at approximately
125\,m intervals\footnote{Six of these strings will make up a dense
subarray in the clearest ice known as Deep Core, extending the
sensitivity of IceCube at lower energies.}. Physics data taking with
IceCube started in 2006 with 9 strings installed. The completion of
the detector construction is planned for the year 2011. The analyses
described here use data taken with the 22-string configuration of the
detector, which operated between 2007 May 31 and 2008 April 5.
 
The data acquisition (DAQ) system of IceCube \citep{nim:a601:294} is
based on local coincidences of photon signals (hits, 
threshold 0.25
photo-electrons) in neighboring or next-to-neighboring DOMs on a
string within $1\,\mu$s. All data from DOMs belonging to a local
coincidence are read out and the digitized waveforms are sent to a
computer farm at the surface. In order to pass the trigger a minimum
number of 8 DOMs in local coincidences within a time window of
$5\,\mu$s is required. If this condition is fulfilled the waveforms
are combined to an event and the number and arrival times of the
Cherenkov photons are extracted. Here, the relative timing resolution
of photons within an event is about 2\,ns. The absolute time of an
event is determined by a GPS clock to a precision of ${\cal
O}({\mu}\mathrm{s})$, which is more than sufficient for our
analyses.

Data are transferred from the South Pole to a computer center in the
North via satellite. For the analyses described in this paper we consider only muons
produced in charged current interactions
\begin{linenomath}\begin{equation}
\nu_{\mu} + N \rightarrow \mu + X 
\end{equation}\end{linenomath}
as only the track-like hit pattern of muons allows for a good angular
resolution. In order to fit satellite bandwidth restrictions, a filter
removes events which do not qualify as good upgoing neutrino
candidates\footnote{All triggered data are stored on tape and shipped
to the North during the Summer season.}. For this purpose, an initial
track is reconstructed for each event using the line-fit algorithm
\citep{nim:a524:169}. This is a simple but fast
analytic track reconstruction based on the measured hit times in the
DOMs. The transferred data forms the basis for our analyses.

\section{Data sets and reconstruction}\label{sec:data}

For our analyses we use data taken with the IceCube detector in its
22-string configuration from 2007 June 1 to 2008 April 4. To prevent
bias in our analyses, the data within the $-1$\,h to $+3$\,h windows
(on-time data) are initially only checked for detector stability until
all parameters of the analysis have been fixed, i.e. they are not used
in the optimization of the analyses. The remaining off-time data that
pass basic quality criteria (95\% of all the data collected in the
22-string configuration) amounts to 268.9 days of livetime. This long
off-time window allows for a precise experimental determination of the
background rate in the on-time windows.

In addition to the basic quality criteria, the on-time data is tested
for stability by fitting a Gaussian to the distribution of low level
event counts in each one second bin. In order to check for unexpected
periods of dead time, an exponential is fit to the distribution of
time delays between events. Data with deviations from the expected
shapes are inspected more closely for possible causes which make the
data unsuitable for our analysis. Here, most observed deviations can
be attributed to transitions between runs and are therefore not
critical. Out of 48 northern hemisphere bursts in the time period
under investigation, the data for seven do not pass the
stability/quality criteria or have gaps during the prompt/precursor
emission windows. For all remaining 41 GRBs, both tests show excellent
agreement with no indications of abnormal behavior of the detector
during the on-time periods. The on-time data cover 100\% of the prompt
and precursor windows for the 41 selected bursts. For the extended
window seven out of the 41 bursts exhibit gaps at the beginning and/or
end of the window. For these seven bursts the extended window is
shortened accordingly (see Table~\ref{tab:timeWindow}). With this
correction, the data cover 94\% of the extended time windows of all
GRBs. The missing 6\% are due to larger gaps in data taking.

\begin{figure}[p]
\epsscale{0.8}
\plotone{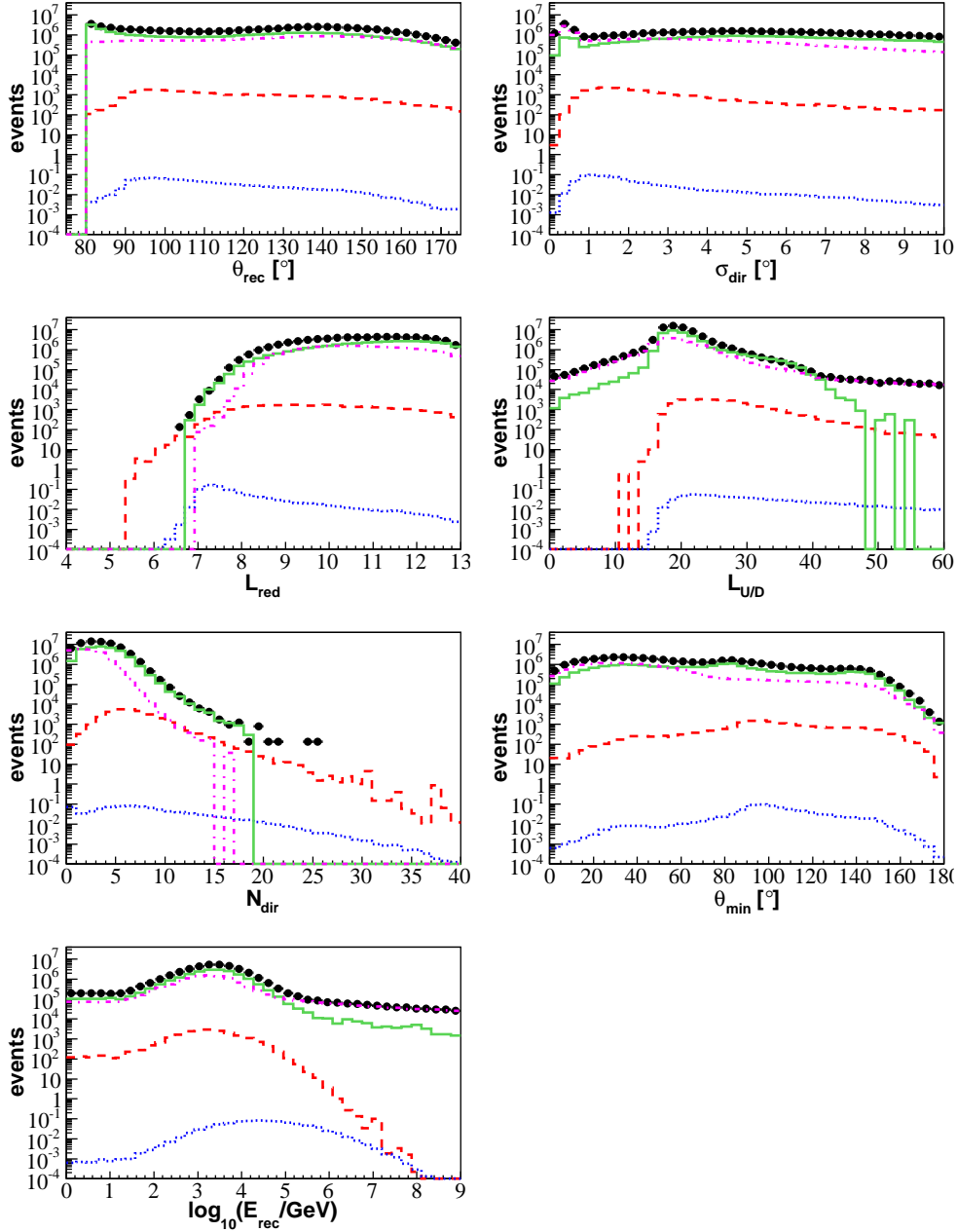}
\caption{Comparison between data (black solid circles; 268.9 days of livetime) and simulations
in the quality parameters used to reject misreconstructed atmospheric
muons at filter level (see Section~\ref{sec:detector}).  Monte Carlo
shown includes atmospheric muons (green solid lines), coincident muons
(magenta dot-dashed lines), atmopheric neutrinos (red dashed lines),
and prompt GRB neutrinos (blue dotted lines).  The GRB signal is
assumed to follow a standard Waxman--Bahcall spectrum and is
normalized to the summed expectation from 41 bursts..
\label{fig:filterLevel}}
\end{figure}

After the data has been transferred to the North, a more precise
determination of the direction of an event is achieved by fitting a
muon-track hypothesis to the hit pattern of the recorded Cherenkov
light in the detector using a log-likelihood reconstruction method
\citep{nim:a524:169}. A fit of a paraboloid to the region around the minimum in
the log-likelihood function yields an estimate of the uncertainty on
the reconstructed direction \citep{app:25:220}. At this point, the
data sample with an event rate of 3.3\,Hz is still dominated by
several orders of magnitude by misreconstructed downgoing atmospheric
muons as demonstrated in Figure~\ref{fig:filterLevel}, which shows a
comparison between data and Monte Carlo (see
Section~\ref{sec:simulation}). The quantities shown are later used to
reject these misreconstructed atmospheric muons and improve the
sensitivity of the analyses.
\tightlist
\item $\theta_\mathrm{rec}$: reconstructed zenith angle\footnote{The
zenith angle in detector coordinates is related to declination
$\delta$ by $\theta = \delta+90^\circ$.};
\item $\sigma_\mathrm{dir}$: the uncertainty on the reconstructed track direction
(quadratic average of the minor and major axis of the $1\sigma$ error
ellipse);
\item $L_\mathrm{red}$: $\log$ of the likelihood value of the reconstructed track
divided by the number of degrees of freedom (number of hit DOMs minus
number of fit parameters). This has proven to be a powerful variable
for separating signal and background as visible from
Figure~\ref{fig:filterLevel};
\item $L_{U/D}$: difference in log-likelihood value between the reconstructed track
and one containing a bias to be reconstructed as downgoing. The bias
is zenith angle dependent and follows the rate of downgoing
atmospheric muons. The rationale behind this is that a track is much
more likely to originate from an atmospheric muon than from a muon
generated in a neutrino interaction. Only high-quality upgoing tracks
have high $L_{U/D}$ values;
\item $N_\mathrm{dir}$: the number of photons detected within a $-15$
to $+75$\,ns time
window with respect to the expected arrival time for unscattered
photons from the muon track hypothesis;
\item $\theta_\mathrm{min}$: minimum zenith angle from a fit of a
two-track hypothesis to the light pattern. For this, the light pattern
is divided into two separate sets of hits based on the mean hit time
of the event, and a track hypothesis is fitted to each hit set
separately. A cut on the smaller zenith angle, $\theta_\mathrm{min}$,
of the two tracks is very effective against so-called \emph{coincident
muons}, where two muons from different atmospheric showers pass the
detector in fast succession mimicking an upgoing track. In this case,
often at least one of the two tracks is reconstructed as downgoing
whereas for good-quality upgoing neutrinos both tracks appear most
often as upgoing; and
\item $E_\mathrm{rec}$: reconstructed muon energy at point of closest
  approach to the center of gravity of hits in an event
  \citep{proc:icrc07:zornoza:1}. This is the calibrated output of a
  likelihood reconstruction method evaluating the measured number of
  photons in each DOM with respect to the corresponding probability
  density function (PDF) of a given track-energy hypothesis.

Muons carry a significant fraction of the original neutrino energy,
and at the energies of highest acceptance in our analyses
(${\sim\!100}\,$TeV), have a range of about 10\,km. The dominant
energy loss mechanisms for these muons are Bremsstrahlung and pair
production which grow with increasing energy, thereby increasing the
amount of Cherenkov light emitted and allowing one to estimate the
muon energy.
\listend

\begin{deluxetable}{lrr}
\tablewidth{0pt}
\tablecaption{Modified Extended Time Window for GRBs with Gaps in Data
Coverage at Beginning or End of $-1$\,h to $+3$\,h
Window\label{tab:timeWindow}}
\tablehead{
\colhead{}
& \colhead{$T_1' - T_0$ [s]}
& \colhead{$T_2' - T_0$ [s]}
}
\startdata
GRB070610  & $-1593$ & $+10800$\\
GRB070714B & $-3600$ &  $+3930$\\
GRB071021  & $-3600$ &  $+5391$\\
GRB071109  & $-3022$ & $+10800$\\
GRB080205  & $-3600$ &  $+6289$\\
GRB080211  & $-3600$ &  $+8399$\\
GRB080320  &  $-389$ &  $+8440$
\enddata
\end{deluxetable}

\begin{deluxetable}{lcccccc}
\tablecaption{Number of Signal and Background (Off-Time Data) Events for the Unbinned Method at Different Cut Levels\label{tab:effUnbinned}}
\tabletypesize{\scriptsize}
\tablewidth{0pt}
%\tablehead{
%   & \multicolumn{2}{c}{Prompt Signal} &
%   \multicolumn{2}{c}{Precursor Signal} & \multicolumn{2}{c}{$E^{-2}$
%   Signal} & \multicolumn{2}{c}{Background} \\ 
%   \colhead{Cut Level} &
%   \colhead{No. Events} & \colhead{$\epsilon$\tablenotemark{a} (\%)} &
%   \colhead{No. Events} & \colhead{$\epsilon$\tablenotemark{a} (\%)} &
%   \colhead{No. Events\tablenotemark{b}} &
%   \colhead{$\epsilon$\tablenotemark{a} (\%)} & \colhead{No. Events} &
%   \colhead{$\epsilon$\tablenotemark{a} (\%)} }
%\startdata
%Filter      &$0.062$	&  100 &$1.8$	&  100 & $7.6$ &  100  & $7.6\times10^8$ & 100 \\
%Final       &$0.033$	&   53 &$0.53$	&   29 & $2.8$	&   37 & $4846$          & $6.5\times10^{-4}$\\
%\enddata
\tablehead{
   & \multicolumn{2}{c}{Prompt Window} &
   \multicolumn{2}{c}{Precursor Window} & \multicolumn{2}{c}{Wide Window}\\
   \colhead{Cut Level} &
   \colhead{No. Events} & \colhead{$\epsilon$\tablenotemark{a} (\%)} &
   \colhead{No. Events} & \colhead{$\epsilon$\tablenotemark{a} (\%)} &
   \colhead{No. Events\tablenotemark{b}} &
   \colhead{$\epsilon$\tablenotemark{a} (\%)} }
\startdata
\cutinhead{Signal}
Filter      &$0.062$	&  100 &$1.8$	&  100 & $7.6$ &  100\\
Final       &$0.033$	&   53 &$0.53$	&   29 & $2.8$	&  37\\
\cutinhead{Background}
Filter\tablenotemark{c}         & $7.6\times10^8$    & 100                & $7.6\times10^8$    & 100                & $7.6\times10^8$ & 100  \\
Final\tablenotemark{c}          & $4846$             & $6.5\times10^{-4}$ & $4846$             & $6.5\times10^{-4}$ & $4846$          & $6.5\times10^{-4}$\\
Final (Window)\tablenotemark{d} & $6.1\times10^{-4}$ & $8.2\times10^{-11}$ & $6.1\times10^{-4}$ & $8.2\times10^{-11}$ & $8.8\times10^{-2}$ & $1.2\times10^{-8}$\\
\enddata
\tablenotetext{a}{Efficiency relative to filter level.} 
\tablenotetext{b}{For a fluence equal to the computed upper limit in Section~\ref{sec:results}.} 
\tablenotetext{c}{All events after cuts.} 
\tablenotetext{d}{Expected events in cones with radii of $2.3^\circ$
(contain 70\% of signal events; see Figure~\ref{fig:cumPSF}) around
GRBs within respective time window (prompt: $T_2-T_1$ from
Table~\ref{tab:grbSpectra}; precursor: 100\,s; wide: 4\,h). Note, that
these cones are \emph{not} used in the evaluation of the data with the
unbinned likelihood method.}
\end{deluxetable}

\section{Simulation}\label{sec:simulation}

Signal neutrinos are generated from the direction of the GRB with the
corresponding spectrum using a port of the \textsc{anis} code
\citep{cpc:172:203} called \textsc{neutrino-generator}. Here, the
change in position of the source in the detector coordinate system
during the respective time window is taken into account. The neutrino
emission is assumed to be constant during this time. In addition,
three types of background are simulated. At the beginning of the
analysis chain, the data sample is dominated by downgoing atmospheric
muons, produced by interactions of cosmic rays in the atmosphere,
which are reconstructed as upgoing. As more cuts are applied the data
sample starts to be dominated by \emph{coincident muon} events. Both event
classes are simulated with the \textsc{corsika} air shower simulation package
\citep{corsika}. Finally, we consider the irreducible background of
atmospheric neutrinos generated in the same interactions as the
atmospheric muons. These neutrinos are simulated as an all-sky flux
with \textsc{neutrino-generator} and weighted according to the Bartol
spectrum \citep{pr:d70:023006}.

Neutrinos are tracked from the surface through the Earth taking into
account absorption, scattering and neutral current regeneration of
neutrinos \citep{cpc:172:203}. Information on the structure of the
Earth is taken from the Preliminary Reference Earth Model
\citep{pepi:25:297}.  Muons originating from neutrino interactions
near the detector and atmospheric muons are traced through rock and
ice taking into account continuous and stochastic energy losses
\citep{hep-ph-0407075}. The photon signal in the DOMs is determined
from a detailed simulation of the propagation of Cherenkov light from
muons and showers through the ice \citep{nim:a581:619} which includes
the modeling of the changes in absorption and scattering length with
depth due to dust layers \citep{jgr:111:D13203}. This is followed by a
simulation of the DOM electronics and the trigger. The simulated DOM
signals are then processed in the same way as the data.

\section{Data analysis}\label{sec:methodology}

Muon neutrinos from GRBs show up as an excess of tracks above the
background from the direction of the GRB within a certain time
window. Background from misreconstructed atmospheric muons can be
suppressed by applying quality cuts on reconstructed
quantities. Background from atmospheric neutrinos on the other hand is
indistinguishable from cosmic neutrinos. Hence, once a high-purity
(atmospheric) neutrino sample has been selected, quality cuts which
aim at rejecting misreconstructed tracks cannot further improve the
signal to background ratio. However, as neutrinos from GRBs are
expected to exhibit a harder energy spectrum than that of atmospheric
neutrinos, information on the muon energy allows to further increase
the sensitivity of the analysis to neutrinos from GRBs. 

We have analyzed the data both with an unbinned likelihood and a
binned method. In order to enhance the chances for a discovery, we do
not analyze the 41 GRBs individually but as a population (stacked
analysis)\footnote{In case of the binned method, the numbers of
expected and observed events of all GRBs are summed up. In case of the
unbinned method, the stacking is performed in Equation
(\ref{eq:grbWeight}).}. This allows us to set the most stringent limit
on the tested models but might not be optimal in other cases, e.g. if
one burst has a much higher neutrino fluence relative to the rest of
the GRBs than expected. Also, in case of a discovery a stacked
analysis only allows one to calculate the neutrino fluence from the
whole burst population but not from individual GRBs. In the following
we describe the unbinned likelihood and binned methods and compare
their performances in the case of the prompt emission scenario. The
more sensitive method is then used to obtain the results presented in
Section~\ref{sec:results}.

\subsection{Unbinned log-likelihood method}

\begin{figure}[p]
\epsscale{0.8}
\plotone{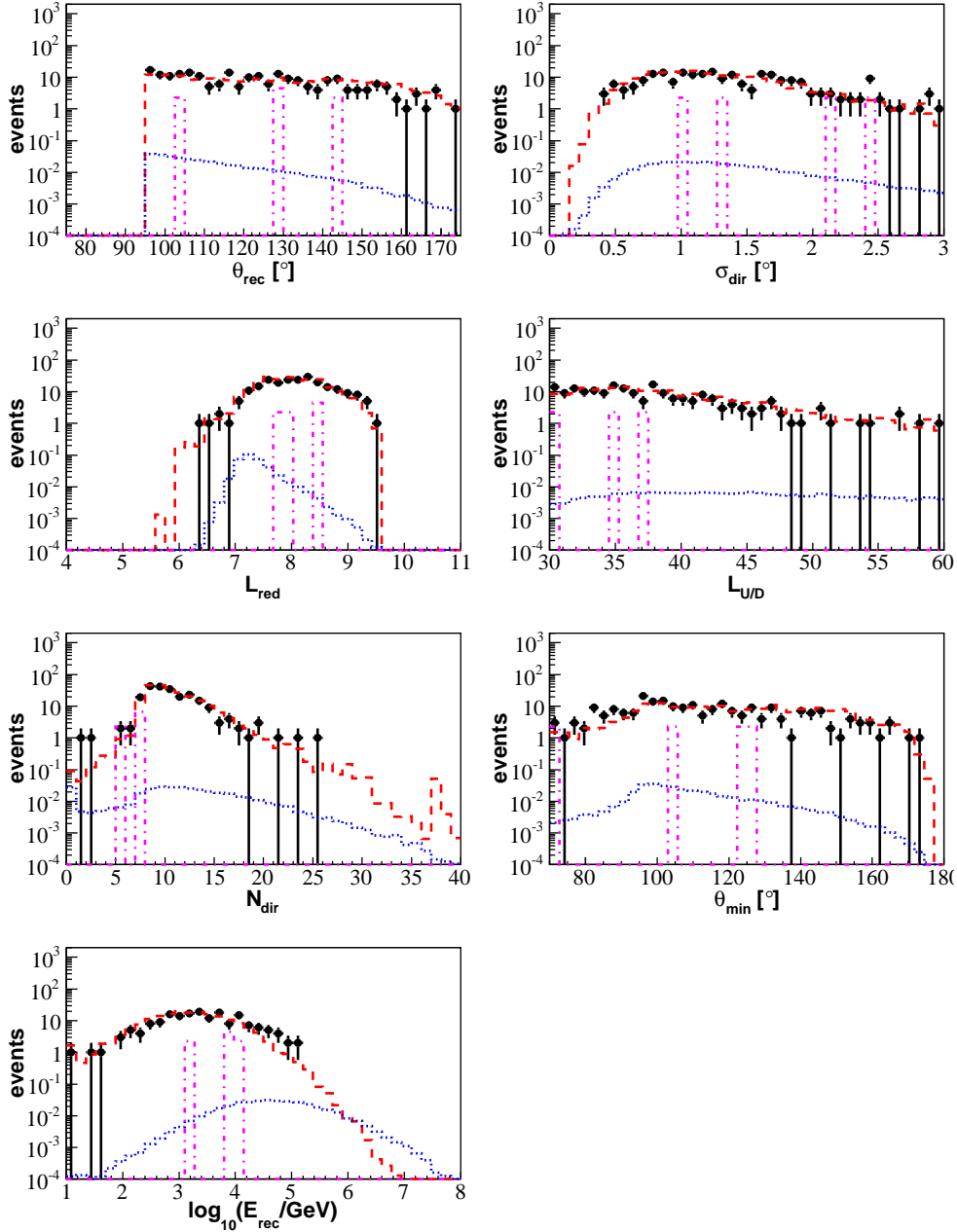}
\caption{Comparison between data (black solid circles; 15.99 days of
livetime) and simulations in the quality parameters used to reject
misreconstructed atmospheric muons at final cut level. 
The data includes all data-taking runs with an overlap with the
($-1$\,h to $+3$\,h) extended search windows (this may be considered
the maximal on-time of the analysis). Monte Carlo
shown includes coincident muons (magenta dot-dashed lines),
atmospheric neutrinos (red dashed lines), and prompt GRB neutrinos
(blue dotted lines). The simulated single atmospheric muons have been
completely removed at this cut level and the statistics for the
simulated coincident muons are very low.  The GRB signal is assumed to
follow a standard Waxman--Bahcall spectrum and is normalized to the
summed contribution of 41 bursts.
\label{fig:dataMCcomp}}
\end{figure}

\begin{figure}[t]
\epsscale{1}
\plottwo{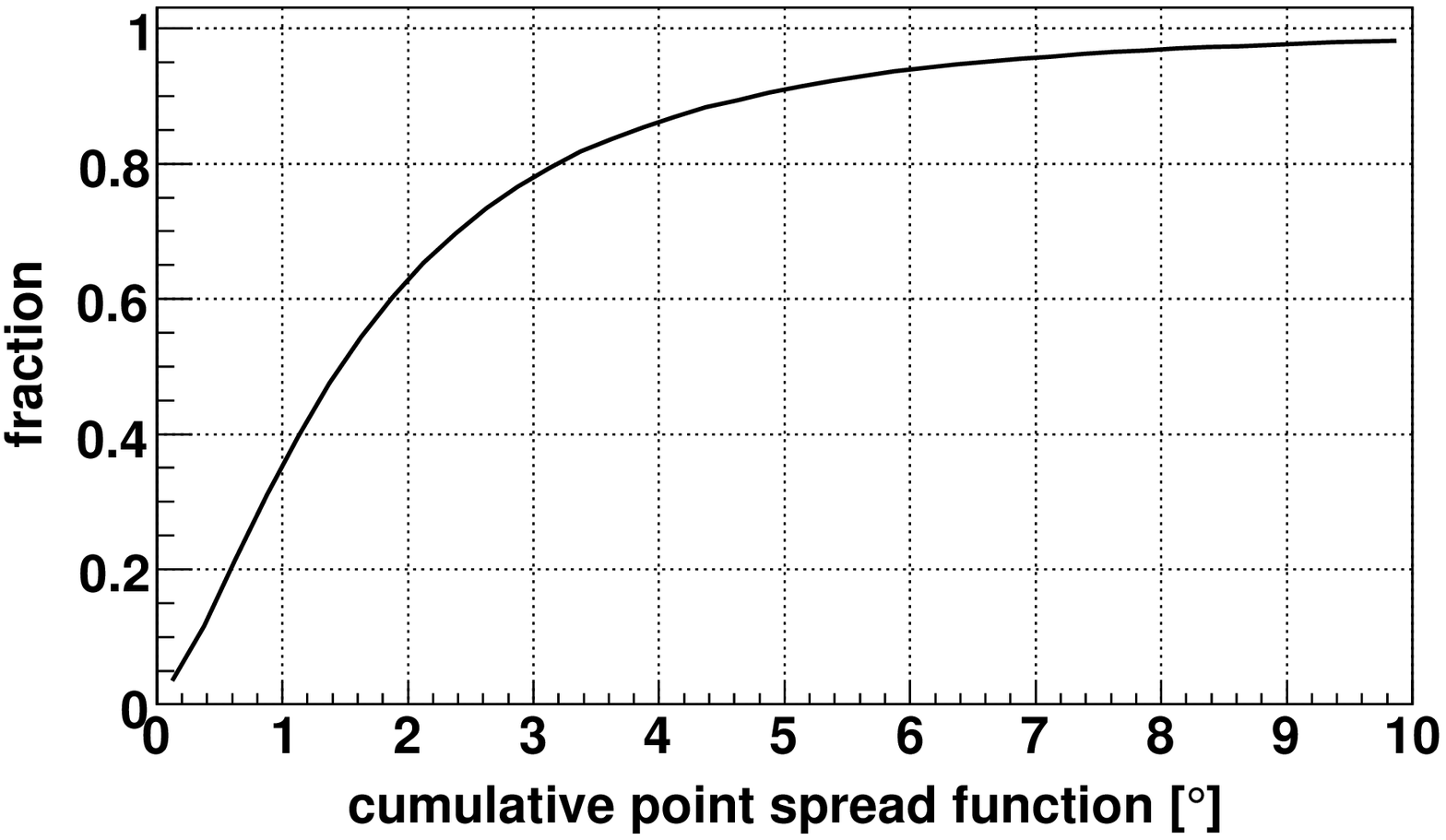}{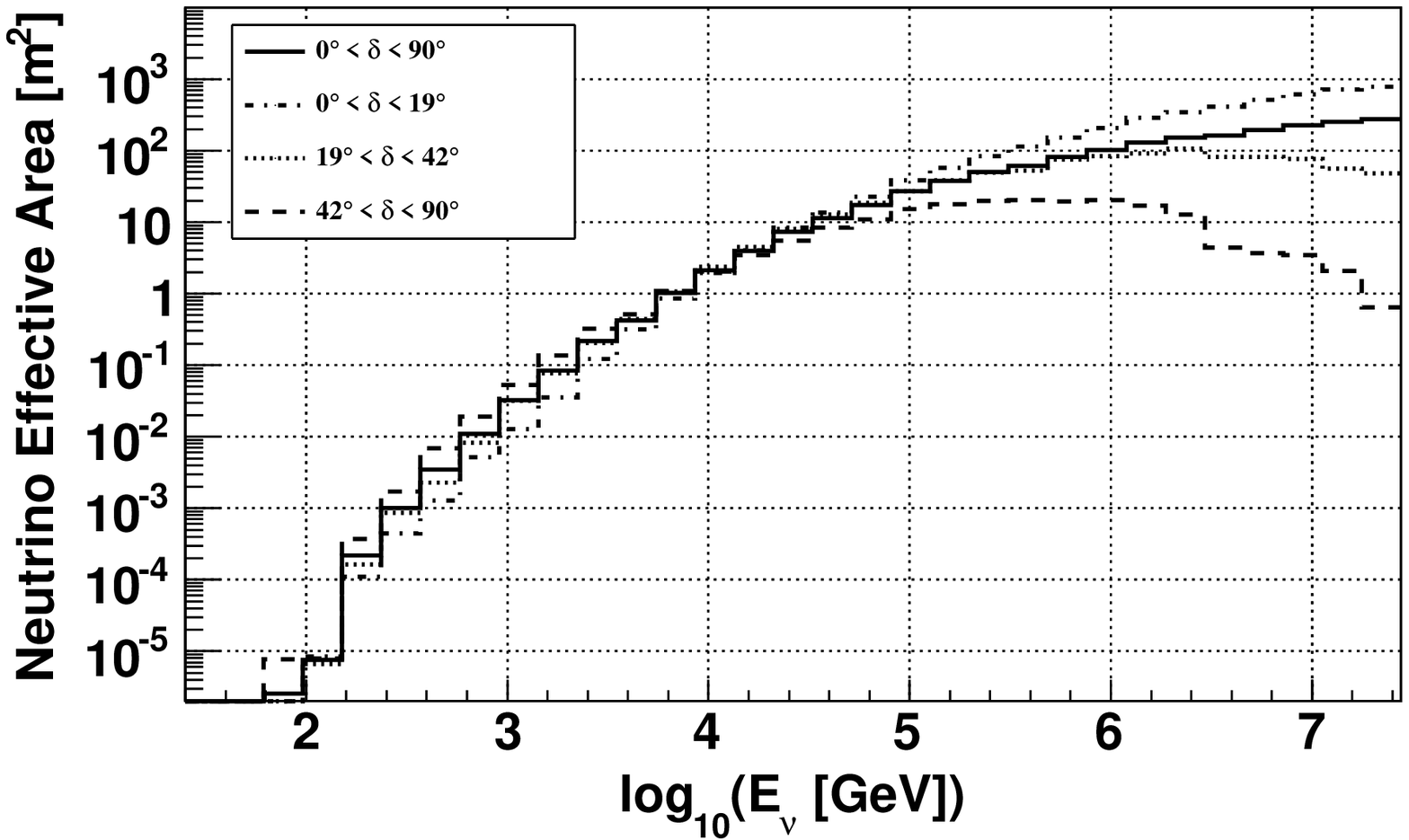}
\caption{Left: Cumulative point spread function for the unbinned
method at final cut level. Right: Effective area for muon 
neutrinos in several declination bands as a function of energy after
final event selection in the unbinned method.\label{fig:cumPSF}}
\end{figure}

After filtering at the South Pole and transfer to the North the
data sample is still dominated by downgoing muons which are
reconstructed as upgoing. These muons are rejected by applying
quality cuts on reconstructed quantities (for a description see
Section~\ref{sec:data}).
\begin{linenomath}\begin{equation}
\theta_\mathrm{rec} > 85^\circ \ \ ; \ \ 
\sigma_\mathrm{dir} < 3^\circ \ \ ; \ \
\theta_\mathrm{min} > 70^\circ \ \ ; \ \
L_{U/D} > 30 \ \ ; \ \
L_\mathrm{red} \le 
  \begin{cases} 
    7.8 \quad \text{for } N_\mathrm{dir} < 7\\
    8.5 \quad \text{for } N_\mathrm{dir} = 7\\
    9.5 \quad \text{for } N_\mathrm{dir} > 7
  \end{cases}.
\label{eq:finalCuts}
\end{equation}\end{linenomath}
After these cuts, a high-purity upgoing (atmospheric) neutrino sample
remains with an event rate of
$2.1\times10^{-4}$\,Hz. Table~\ref{tab:effUnbinned} lists the cut
efficiency for data and signal. A comparison of data with simulations
is displayed in Figure~\ref{fig:dataMCcomp}. In general, good agreement
between the atmospheric neutrino Monte Carlo and data is
observed. Small deviations for example visible at low $N_\mathrm{dir}$
are most likely due to atmospheric muon background which is not
accounted for by the Monte Carlo due to its limited statistics at this
cut level. The cumulative point spread function is shown in
Figure~\ref{fig:cumPSF} (left).
The median angular resolution is about
$1.5^\circ$. Figure~\ref{fig:cumPSF} (right) shows the muon neutrino
effective area for different declination bands.

The data sets after quality cuts are the starting point for the
unbinned likelihood method. In contrast to binned methods where the
event is rejected if it lies outside the cut region (binary
selection), unbinned likelihood methods do not discard events but use
PDFs to evaluate the probability of an event belonging to signal or
background population. The unbinned likelihood method used here is
similar to that described in \citet{app:29:299}. The signal,
$S(\vec{x}_i)$, and background, $B(\vec{x}_i)$, PDFs are each the
product of a time PDF, a directional PDF, and an energy PDF, where
$\vec{x}_i$ denotes the directional, time, and energy variables.

\begin{figure}[t]
\epsscale{0.6}
\plotone{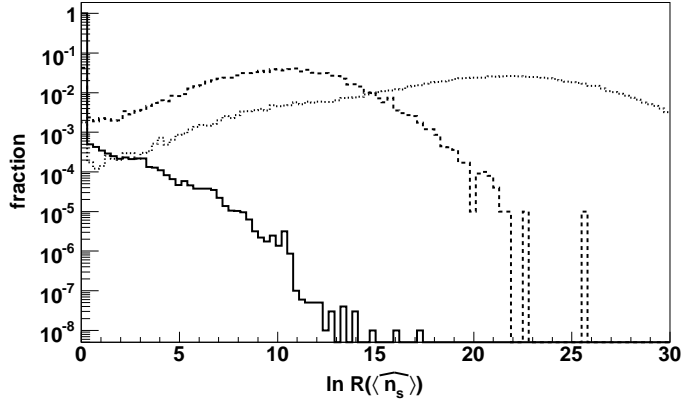}
\caption{Likelihood-ratio distribution of $10^8$ randomized background-only data
sets for the prompt-window analysis (solid line). Also shown are the
corresponding distributions for background data sets with one (dotted)
and two (fine dotted) signal events injected. The integrals of all
distributions have been normalized to one.\label{fig:nullHisto}}
\end{figure}

The directional signal PDF is a two-dimensional Gaussian distribution
with the two widths being the major and minor axes of the $1\sigma$
error ellipse of the paraboloid fit described in
Section~\ref{sec:data}. The time PDF is flat over the respective time
window and falls off on both sides with a Gaussian distribution. The
width $\sigma$ of the Gaussian is determined by the length of the time
window with a maximum of $\sigma = 25$\,s and a minimum of $\sigma =
2$\,s. The Gaussian accounts for possible small shifts in the neutrino
emission time with respect to that of the $\gamma$-rays and prevents
discontinuities in the likelihood function. The sensitivity of the
method depends only weakly on the exact choice of $\sigma$. The
energy PDF is determined for each GRB individually. It is derived from
the energy-estimator distribution of the tracks of the corresponding
signal Monte Carlo data set (weighted to an $E^{-2}$
spectrum\footnote{Using an energy PDF different from that of the
assumed signal spectrum cannot lead to an overestimation of the
significance of a potential signal as the same PDF is also used to
obtain the distribution of background-only samples (see
Figure~\ref{fig:nullHisto}) from which the significance is
calculated. Studies have shown that for our analyses the loss in
discovery potential due to the different PDFs is very small. On the
other hand, the softer energy PDF increases the sensitivity to
scenarios where the true GRB spectra are softer than assumed.}) after
final cuts (see Equation (\ref{eq:finalCuts})). The signal PDFs of the
GRBs are combined using a weighted sum
\citep{apj:636:680}
\begin{linenomath}\begin{equation}
S_\mathrm{tot}(\vec{x}_i) = \frac{\sum_{j=1}^{N_\mathrm{GRBs}}
w_j\,S_j(\vec{x}_i)}{\sum_{j=1}^{N_\mathrm{GRBs}} w_j} \ ,
\label{eq:grbWeight}
\end{equation}\end{linenomath}
where $S_j(\vec{x}_i)$ is the signal PDF of the $j$th GRB and $w_j$ is a
weight that in the case of the prompt and precursor window is
proportional to the expected number of events in the detector
according to the fluences described in Section~\ref{sec:speccalc}. In
the case of the extended window we use $w_j = 1$ for all GRBs in
order to make the search as general as possible.

For the directional background PDF the detector asymmetries in zenith
and azimuth must be taken into account. This is accomplished by
evaluating the data in the detector coordinate system. The directional
background PDF is hence derived from the distribution of all off-time
events after final event selection in the zenith--azimuth plane of the
detector. The time distribution of the background during a GRB can be
assumed to be constant resulting in a flat time PDF. The energy PDF is
determined in the same way as for the signal PDF with weights
corresponding to the Bartol atmospheric neutrino flux.

All PDFs are combined in an extended log-likelihood function
\citep{barlow:statistics:1989}
\begin{linenomath}\begin{equation}
\ln\left({\cal L}(\langle n_s \rangle)\right) = -\langle n_s \rangle - \langle n_b \rangle + \sum_{i=1}^{N} \ln\left(
\langle n_s \rangle\,S_\mathrm{tot}(\vec{x}_i) + \langle n_b \rangle\,B(\vec{x}_i) \right) \ ,
\end{equation}\end{linenomath}
where the sum runs over all reconstructed tracks in the final sample. The
variable $\langle n_b \rangle$ is the expected mean number of
background events, which is determined from the off-time data
set. The mean number of signal events, $\langle n_s \rangle$, is a
free parameter which is varied to maximize the expression
\begin{linenomath}\begin{equation}
\ln\left({\cal R}(\langle n_s \rangle)\right) = \ln\left(\frac{{\cal L}(\langle n_s \rangle)}{{\cal L}(0)}\right) = -\langle n_s \rangle +
	\sum_{i=1}^N \ln \left(
	\frac{\langle n_s \rangle\,S_\mathrm{tot}(\vec{x}_i)}{\langle n_b \rangle\,B(\vec{x}_i)} + 1 \right)
\label{eq:lnR-dist}
\end{equation}\end{linenomath}
in order to obtain the best estimate for the mean number of signal
events, $\widehat{\langle n_s \rangle}$.

To determine whether a given data set is compatible with the
background-only hypothesis $10^8$ background data sets for the on-time
windows are generated from off-time data by randomizing the track
times while taking into account the downtime of the detector. For each
of these data sets the $\ln({\cal R})$ value is calculated, yielding
the distribution shown in Figure~\ref{fig:nullHisto}. The probability
for a data set to be compatible with background is given by the
fraction of background data sets with a larger $\ln({\cal R})$
value. For comparison, the plot also displays the $\ln({\cal R})$
distributions for background data sets with one and two injected Monte
Carlo signal events, respectively. The signal events are randomly
distributed among the GRBs, where the assignment probability to a
specific GRB is proportional to the expected number of events from
that burst. The energy of the neutrinos is generated according to the
spectra calculated in Section~\ref{sec:nuFluxesPrompt}.

\subsection{Binned method}

For the binned method, a machine learning algorithm was trained to
separate signal and background. The algorithm used was a Support
Vector Machine (SVM) \citep{ml:20:273} with a radial basis function
kernel.  It was provided with the best reconstructed track direction
in detector coordinates as well as many quality parameters including
$\sigma_\mathrm{dir}$, $L_\mathrm{red}$, $L_{U/D}$, $N_\mathrm{dir}$,
and $\theta_\mathrm{min}$ which are described in
Section~\ref{sec:data}.  The SVM was trained using the off-time
filtered data as background and all-sky neutrino simulation weighted
to the sum of the individual burst spectra as signal.  The optimum SVM
parameters (kernel parameter, cost factor, margin) were determined
using a coarse, and then fine, grid search with a 5-fold cross
validation technique at each node, as described in
\citet{svm:practical}.

\begin{figure}[t]
\epsscale{0.5}
\plotone{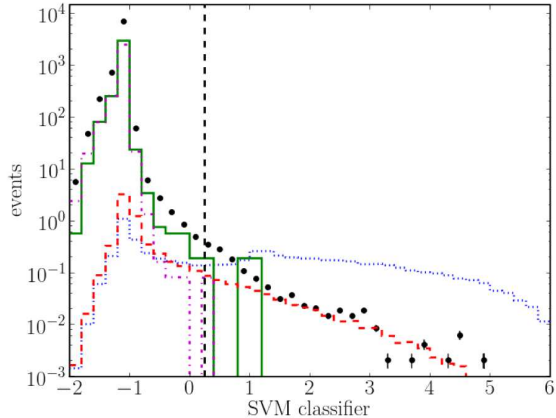}
\caption{The SVM classifier distribution of data (black solid
  circles) and simulations.  Monte Carlo shown includes atmospheric muons
  (green solid lines), coincident muons (magenta dot-dashed lines),
  atmospheric neutrinos (red dashed lines), and prompt GRB neutrinos (blue 
  dotted lines).  The GRB signal is assumed to follow the summed calculated 
  individual neutrino spectra and is normalized to the rate of atmospheric neutrinos.
  The vertical dashed line indicates the final optimum cut at 0.25. \label{fig:binned}}
\end{figure}

The resulting SVM classification of events is shown in
Figure~\ref{fig:binned}. The final cut on this classifier is optimized to
detect a signal fluence with at least 5$\sigma$ (significance) in 50\%
of cases (power) by minimizing the Model Discovery Factor (MDF)
according to
\citet{proc:phystat05:hill:1}. The MDF is the ratio between the signal
fluence required for a detection with the specified significance and
power and the predicted fluence. The angular cut around each GRB is
then calculated to keep 3/4 of the remaining signal after the cut on
the SVM classifier. In this way, there is one cut on the SVM classifier
for all GRBs, but different angular cuts around each GRB according to
the angular resolution of the detector in that direction. The optimum SVM cut
is determined to be at a value of 0.25.
Table~\ref{tab:effBinned} displays the signal and background event
rates at different cut levels.

\begin{deluxetable}{lcccc}
\tablecaption{Number of Signal and Background (Off-Time Data) Events
  for the Binned Method in the Prompt Window at Different Cut Levels\label{tab:effBinned}}
\tabletypesize{\scriptsize}
\tablewidth{0pt}
\tablehead{
                    & \multicolumn{2}{c}{Signal} &
                    \multicolumn{2}{c}{Background} \\
\colhead{Cut Level} & \colhead{No. Events} & \colhead{Efficiency\tablenotemark{a} (\%)} & \colhead{No. Events} & \colhead{Efficiency\tablenotemark{a} (\%)} 
}
\startdata
Filter      &$0.062$	&  100 & $77\times10^7$  & 100 \\
Final       &$0.023$	&   37 & $4.7$  & $6.1\times10^{-9}$ \\
\enddata
\tablenotetext{a}{Relative to filter level.}
\end{deluxetable}

\subsection{Comparison of the two methods} 
\begin{figure}[t]
\epsscale{0.6}
\plotone{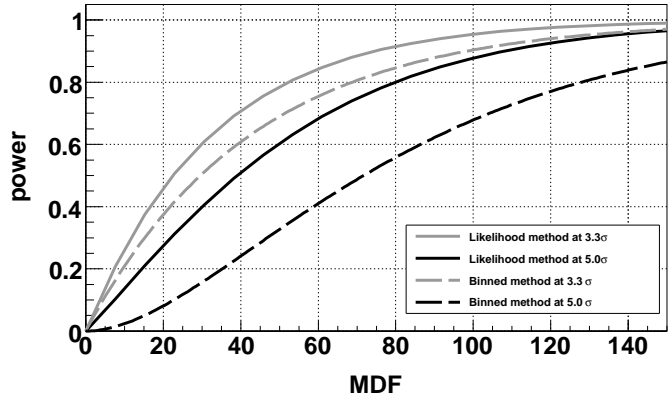}
\caption{Comparison of the discovery potentials of the unbinned
likelihood (solid) and binned (dashed) method for significances of
$3.3\sigma$ (light) and $5\sigma$ (dark). Shown is the fraction of
data sets yielding at least the stated significance (power) as a
function of the ratio between the mean number of injected signal
events and the expected number of signal events from the model
(MDF).\label{fig:discPotential}}
\end{figure}

We compare the performance of the unbinned likelihood and binned
method by means of their discovery potential for the \emph{prompt}
neutrino emission scenario.  Figure~\ref{fig:discPotential} displays
for a significance of $5\sigma$ the power as a function of the MDF for
the two methods. For a power of 50\% the unbinned method
shows an improvement in the MDF of about a factor 1.8 compared to the
binned method.  It is therefore used to derive the results presented
in the following section. The large gain in sensitivity is partly due
to the explicit use of energy information in the likelihood. The
discovery potential for the expected fluxes is further improved by
weighting the bursts according to the expected number of signal events
in the detector in the unbinned method (Equation
(\ref{eq:grbWeight})). In the binned method all bursts are treated
equally.

\section{Results and systematic uncertainties}\label{sec:results}

\begin{figure}[t]
\epsscale{.65}
\plotone{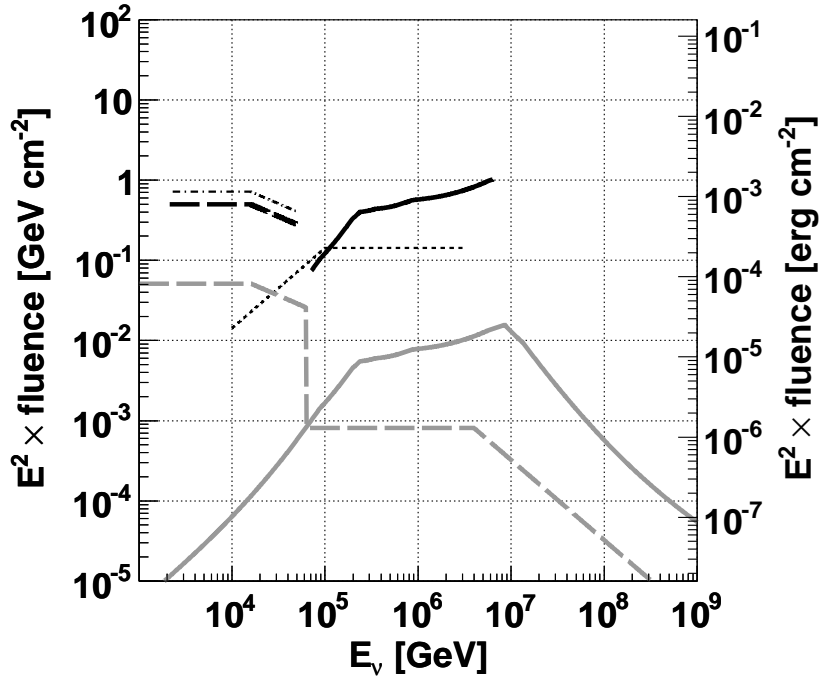}
\caption{90\% CL upper limits (dark thick lines) on the neutrino fluence from
the 41 northern hemisphere GRBs for different emission models (light
thick lines): precursor (dashed, \citet{pr:d68:083001}) and prompt
(solid, see Section~\ref{sec:speccalc}). The dark dotted and
dash-dotted thin lines mark the scaled AMANDA 90\% CL upper limits
\citep{apj:674:357,kuehn:2007} on the prompt and precursor fluences,
respectively.  \label{fig:limits}}
\end{figure}

We apply the unbinned likelihood method to the on-time data sets after
neutrino candidate event selection with the final cuts (Equation
(\ref{eq:finalCuts})). For all three emission scenarios the values of
$\ln({\cal R})$ and $\widehat{\langle n_s \rangle}$ are zero and hence
consistent with the null hypothesis.Therefore, we derive 90\% CL upper
limits\footnote{The limits are calculated following a procedure due to
  Neyman \citep{ptrsl:a236:333,pl:b667:1}. The 90\% CL upper limit
  corresponds to the signal flux for which 90\% of background data
  sets with signal events injected according to the calculated GRB
  spectra yield $\ln({\cal R})$ values greater than the observed one.}
on the fluence from the 41 GRBs in the prompt phase of $3.7 \times
10^{-3}\,\mathrm{erg}\,\mathrm{cm}^{-2}$ (72\,TeV -- 6.5\,PeV) and on
the fluence from the precursor phase of $2.3 \times
10^{-3}\,\mathrm{erg}\,\mathrm{cm}^{-2}$ (2.2\,TeV -- 55\,TeV), where
the quoted energy ranges contain 90\% of the expected signal events in
the detector. Further information is listed in
Table~\ref{tab:results}. The limits, which are displayed in
Figure~\ref{fig:limits}, are not strong enough to constrain the
models. The 90\% CL upper limit for the wide time window is $2.7
\times 10^{-3}\,\mathrm{erg}\,\mathrm{cm}^{-2}$ (3\,TeV -- 2.8\,PeV)
assuming an $E^{-2}$ flux. 

To illustrate, we counted the number of events after final
cuts in cones with radii $2.3^\circ$ around the GRB positions (contain
70\% of signal events; see Figure~\ref{fig:cumPSF}) within the
corresponding time windows (prompt: $T_2-T_1$ from
Table~\ref{tab:grbSpectra} ; precursor: 100\,s; wide: 4\,h). In
addition, we analyzed the data in the prompt window with the
binned method after final cuts. In all cases, zero events remain which
is consistent with the results of the unbinned likelihood method.

\begin{deluxetable}{lccc}
\tablewidth{0pt}
\tablecaption{Summary of Search Results for Prompt and Precursor Window\label{tab:results}}
\tablehead{
\colhead{Window}
& \colhead{$n_\mathrm{exp}$}
& \colhead{$n_\mathrm{limit}$}
& \colhead{Factor}
}
\startdata
Prompt Window    & 0.033 & 2.4 & 72\\
Precursor Window & 0.26  & 2.5 & 9.7\\
Wide Time Window & --    & 2.7 & --\\
\enddata
\tablecomments{$n_\mathrm{exp}$: number of expected events in the
detector after final cuts from all 41 GRBs in the unbinned search;
$n_\mathrm{limit}$: 90\% CL upper limit on the event number from all
41 GRBs; factor: factor by which the limit exceeds the predicted event number $n_\mathrm{exp}$.}
\end{deluxetable}

As described previously, we use the off-time data to determine the
background rate in the on-time windows. This technique removes many
potential sources of uncertainties in the calculation of the
significance of a possible signal that are introduced when using a
simulation of the background. However, this method makes the
assumption that the rate of data during the off-time and on-time
windows are the same. Furthermore, we use Monte Carlo for the signal
simulation and the derivation of upper limits, which involves the
propagation of particles through the Earth and ice, and the simulation
of the detector response. The most important sources for systematic
uncertainties are discussed below in detail. Their effects on the
upper limits are summarized in Table~\ref{tab:sysErr}.
\begin{deluxetable}{lccc}
\tablewidth{0pt}
\tablecaption{Summary of Effects of Systematic Uncertainties on the
Upper Limits\label{tab:sysErr}}
\tablehead{
\colhead{Type of Uncertainty}
& \colhead{Prompt Window}
& \colhead{Precursor Window}
& \colhead{Extended Window}
}
\startdata
Ice Simulation     & $\pm15\%$ & $\pm15\%$  & $\pm15\%$ \\
DOM Efficiency     & $\pm 5\%$ & $\pm 10\%$ & $\pm 7\%$ \\
Lepton Propagation & $\pm 5\%$ &  $\pm 5\%$ & $\pm 5\%$ \\
Background Rate    &    $<1\%$ &     $<1\%$ &    $<1\%$ \\
\hline
Sum                & $\pm17\%$ &  $\pm19\%$ & $\pm17\%$ \\
\enddata
\end{deluxetable}

\emph{Ice simulation:} Inaccuracies in the ice simulation can lead to
a wrong estimate of the efficiency of the detector to neutrinos from
GRBs. Data-Monte Carlo comparisons and variation of simulation
parameters indicate that the systematic uncertainty from this aspect
of the simulation is about $\pm 15\%$.

\emph{DOM efficiency:} A $\pm10\%$ uncertainty in the efficiency of
the optical modules in the detection of photons leads to a
corresponding uncertainty in the number of expected events from a GRB.
This effect is nonlinear and spectrally dependent, so simulation was
generated spanning the range of uncertainties to determine the
resulting change in signal event rates;

\emph{Neutrino and muon propagation:} Theoretical uncertainties on muon
energy losses and the neutrino-nucleon cross-section, determined from
the uncertainty on the CTEQ6 PDFs \citep{jhep:0207:012},
contribute a 5\% uncertainty on the neutrino event rate in the detector;

\emph{Background rate:} After final cuts the variation of the event
rate over the data taking period is about $\pm5\%$. In order to
account for potential differences at the time of the bursts the
background data rate is varied by this amount. This results in a
shift of the upper limits of less than $\pm1\%$ and is therefore
negligible.

\section{Comparison to other results}\label{sec:otherResults} 
A search with the AMANDA detector for muon neutrinos in the
prompt and precursor phase was conducted for GRBs detected between
1997 and 2003 \citep{apj:674:357} with null result. The analysis of
prompt neutrinos contained 419 bursts observed by the BATSE experiment
\citep{apjs:122:465} as well as by others with similar
characteristics. The large number of bursts allowed that analysis to
set an upper limit only a factor 1.4 above the prediction of the
Waxman--Bahcall prompt emission model. A further search with the AMANDA
detector for muon neutrinos was conducted for 85 GRBs detected between
2005 and 2006 \citep{strahler:2009}, primarily by the \emph{Swift} satellite,
also with null result. Due to the smaller number of bursts, the upper
limit from this analysis is much less restrictive. Converting the
upper limit obtained from the 419 bursts to a fluence limit
from 41 standard Waxman--Bahcall bursts (for definition see footnote
\ref{foot:wbburst}) yields the dotted line in
Figure~\ref{fig:limits}. Due to the 10 times smaller number of bursts
available, the limit presented in this paper is about a factor three
worse. For the precursor emission model, the AMANDA limit
\citep{apj:674:357,kuehn:2007} is much less restrictive as only 60
bursts, detected between 2001 and 2003, were used. It is shown in
Figure~\ref{fig:limits} as a dash-dotted line. Here, our analysis
improves on the AMANDA upper limit despite the fact that 30\% fewer
bursts were investigated.

The AMANDA data were also analyzed for neutrinos of all flavors from
GRBs \citep{apj:664:397}. Apart from a search for neutrinos from 73
bursts detected by BATSE in 2000, another search did not rely on
information from satellites but looked for a clustering of events
within sliding time windows of 1\,s and 100\,s. Due to this more
generic approach and the low number of bursts, respectively, the
limits from these analyses are much less restrictive than those from
\citet{apj:674:357}. However, the sliding-window or similar searches
are the only way to detect GRBs where either the jet does not emerge
from the progenitor star (choked bursts,
\citet{prl:87:171102}) or the $\gamma$-ray signal is not observed by
satellites for other reasons.

We continue with a discussion of the impact of results from
high-energy $\gamma$-ray observations on expected neutrino fluences.
Along with high-energy neutrinos which originate from the decay of
charged pions, high-energy $\gamma$-ray photons are produced in the decay
of simultaneously generated neutral pions. In addition, high-energy
photons are produced in inverse-Compton scattering of synchrotron
photons by accelerated electrons
\citep{arXiv:0810.0520}. In contrast to neutrinos, the flux of
high-energy $\gamma$-ray photons at the Earth is significantly reduced due
to the large optical depths for photon-photon pair production inside
the source for not too large jet Lorentz factors $\Gamma_\mathrm{jet}
\lesssim 800$ \citep{arXiv:0810.0520}. In addition, high-energy photons above
100\,GeV are absorbed on the extragalactic background light (EBL) if they
travel distances with $z \gtrsim 0.5$. Observations with air-Cherenkov
telescopes like H.E.S.S.\ \citep{apj:690:1068,arXiv:0902.1561} or MAGIC
\citep{apj:667:358,proc:aip:magic:1} are also hampered by the fact
that usually it takes more than 50\,s (MAGIC) or 100\,s (H.E.S.S.)
from the observation of a GRB by a satellite to the start of data
taking with these telescopes. Therefore, the prompt emission window is
only partially covered or not at all. MILAGRO as an air shower array
observed large parts of the sky continuously
\citep{apj:604:L25,apj:630:996,apj:666:361}. However, it was mostly
sensitive to energies above 100\,GeV and therefore suffered
significantly from $\gamma$-ray absorption on the EBL. HAWC, the
successor of Milagro currently in the planning phase, will be able to
detect $\gamma$ rays from GRBs down to 100\,GeV where about 50
(\emph{Fermi}) GRBs per year will fall in its field of view
\citep{proc:icrc09:goodman:1}.

At energies above 100\,GeV, there has been no definitive detection of
$\gamma$-ray emission from GRBs. Milagrito
\citep{apj:533:l119,apj:583:824} and the HEGRA AIROBICC array
\citep{aa:337:43} reported evidence at the $3\sigma$ level for
high-energy $\gamma$-ray emission from GRB\,970417A ($E_\gamma >
650$\,GeV) and GRB\,920925C ($E_\gamma > 20$\,TeV), respectively.
However, subsequent searches for high-energy $\gamma$-ray emission
from GRBs did not find similar signals. The limits obtained from MAGIC
and H.E.S.S.\ are not directly comparable to our results due to their
incomplete burst coverage. In \citet{apj:604:L25}, \citet{apj:630:996}
and \citet{apj:666:361} the MILAGRO collaboration reports 99\% CL
upper limits for a large number of individual bursts (both long and
short) between 2000 and 2006 down to
$10^{-7}\,\mathrm{erg}\,\mathrm{cm}^{-2}$ (energy range
$\sim\!\!100$\,GeV -- 10\,TeV; the exact energy range differs from
publication to publication). Extrapolating the average per burst 99\%
CL upper limit for the prompt window in our analysis to the energy
range from 100\,GeV to 10\,TeV yields $3.2\times10^{-7}
\,\mathrm{erg}\,\mathrm{cm}^{-2}$. However, the photon limits do not
constrain our results as they do not account for absorption in the EBL
(this would significantly worsen the limits) and include an unknown,
probably dominant, contribution from inverse-Compton scattering. In
general, current flux predictions for high-energy gamma rays from GRBs
are near or below the sensitivity of current instruments
\citep{arXiv:0810.0520}, where the predicted fluxes in the energy
range below ${\sim\!100}$\,TeV are dominated by the leptonic emission
component in most scenarios.

Within the internal shock (fireball) model, synchrotron self-Compton
(SSC) processes between the accelerated electrons and the $\gamma$-ray
photons could lead to a high-energy $\gamma$-ray peak in the GeV range
detectable by \emph{Fermi} in case of bright GRBs. Up to now, this has
happend only for a handful of bursts (one of the photons with the
highest energy from a GRB detected by \emph{Fermi} came from
GRB\,080916C and had an energy of $\sim\!13$\,GeV
\citep{science:323:1688}). However, this does not necessarily disfavor
the internal shock scenario. As discussed in \citet{mnras:sub:1}, the
amount of energy in high-energy photons could be suppressed by an
inefficient SSC process in the extreme Klein-Nishina regime or a
combination of a SSC peak at high energies and a low photon cut-off
energy above which the fireball becomes optical thick.

\section{Conclusions and outlook}

We have performed a set of complementary searches for muon neutrinos
associated in space and time with 41 gamma-ray bursts that were
observed in the northern sky between 2007 June and 2008 April.  For
the first time in searches with large GRB populations, we have
calculated individual prompt neutrino spectra for all 41 GRBs using
measured GRB parameters. The search results are consistent with the
case of a background-only hypothesis. Therefore, we place 90\% CL
upper limits on the fluence from the prompt phase of $3.7 \times
10^{-3}\,\mathrm{erg}\,\mathrm{cm}^{-2}$ (72\,TeV -- 6.5\,PeV) and on
the fluence from the precursor phase of $2.3 \times
10^{-3}\,\mathrm{erg}\,\mathrm{cm}^{-2}$ (2.2\,TeV -- 55\,TeV), where
the quoted energy ranges contain 90\% of the expected signal events in
the detector. Though the number of bursts is smaller than in previous
searches the larger detector allows us to improve on the limits for
the precursor phase by a factor 1.4. Compared to the predictions, the
limits lie a factor 72 (prompt phase) and 9.7 (precursor phase)
higher. Hence, they do not allow us to constrain the models. Apart
from these model-driven searches, we have also conducted for the first
time a generic search for neutrino emission from GRBs in a wide window
of ($-1$\,h to $+3$\,h) around each burst. Finding no evidence for a
signal, we place a 90\% CL upper limit on the fluence of $2.7 \times
10^{-3}\,\mathrm{erg}\,\mathrm{cm}^{-2}$ (3\,TeV -- 2.8\,PeV) assuming
an $E^{-2}$ flux.

Launched in 2008 June, the \emph{Fermi} Gamma-ray Space Telescope
\citep{fermi:homepage} has begun to provide an expanded catalog of
sources for future neutrino searches. With a much larger field of view
than other satellites, \emph{Fermi} has increased the GRB detection
rate by more than a factor two. At the same time, the detected bursts
have on average a higher luminosity than those detected by
\emph{Swift} due to the lower sensitivity of the \emph{Fermi}-GBM
(Gamma-ray Burst Monitor) instrument. The first IceCube analysis to
take advantage of the increased detection opportunities will utilize
the 40-string configuration of the detector, already as large as the
full IceCube along one axis. This gives it the full angular resolution
power along that direction and thus provides powerful background
rejection.  However, this is mitigated by the comparatively poor
angular resolution of the \emph{Fermi}-GBM ($\sim3^{\circ}$), which is
worse than the IceCube resolution. With the full 80-string detector
scheduled to be completed in 2011 and an expected 100 to 150 detected
bursts per year in the northern hemisphere by the
\emph{Fermi} and \emph{Swift} satellites the sensitivity of IceCube to
neutrinos from GRBs will soon exceed that of AMANDA. This will allow
IceCube to either confirm the predicted fluxes within the next years
or set stringent limits thereby disfavoring GRBs as the major sources
of ultra-high energy cosmic rays.

%% IceCube Acknowledgements %%

\acknowledgments

We acknowledge the support from the following agencies:
U.S. National Science Foundation-Office of Polar Program,
U.S. National Science Foundation-Physics Division,
University of Wisconsin Alumni Research Foundation,
U.S. Department of Energy, and National Energy Research Scientific Computing Center,
the Louisiana Optical Network Initiative (LONI) grid computing resources;
Swedish Research Council,
Swedish Polar Research Secretariat,
and Knut and Alice Wallenberg Foundation, Sweden;
German Ministry for Education and Research (BMBF),
Deutsche Forschungsgemeinschaft (DFG), Germany;
Fund for Scientific Research (FNRS-FWO),
Flanders Institute to encourage scientific and technological research in industry (IWT),
Belgian Federal Science Policy Office (Belspo);
the Netherlands Organisation for Scientific Research (NWO);
Marsden Fund, New Zealand;
M.~Ribordy acknowledges the support of the SNF (Switzerland);
A.~Kappes and A.~Gro{\ss} acknowledge support by the EU Marie Curie OIF Program;
J.~P.~Rodrigues acknowledge support by the Capes Foundation, Ministry of Education of Brazil.

\appendix
\section{Equations used in the calculation of the neutrino spectra}\label{app:nu_spec}
\begin{linenomath}\begin{align}
F_\gamma(E_\gamma) &= \frac{\mathrm{d}N(E_\gamma)}{\mathrm{d}E_\gamma} = f_\gamma \times
             \begin{cases}
             \left(\frac{\epsilon_\gamma}{\mathrm{MeV}}\right)^{\alpha_\gamma} \ \left(\frac{E_\gamma}{\mathrm{MeV}}\right)^{-\alpha_\gamma} 
                & \text{for $E_\gamma < \epsilon_\gamma$} \\
             \left(\frac{\epsilon_\gamma}{\mathrm{MeV}}\right)^{\beta_\gamma} \  \left(\frac{E_\gamma}{\mathrm{MeV}}\right)^{-\beta_\gamma} & \text{for $E_\gamma \ge\epsilon_\gamma$}
             \end{cases}
\label{eq:photonspec}\\[2mm]
{\cal F}_\gamma &=
		\int \mathrm{d}E_\gamma
                \ E_\gamma F_\gamma(E_\gamma) \label{eq:gammaIntegral}\\[5mm]
F_\nu(E_\nu) &= \frac{\mathrm{d}N(E_\nu)}{\mathrm{d}E_\nu} = f_\nu \times
             \begin{cases}
             \left(\frac{\epsilon_{\nu,1}}{\mathrm{GeV}}\right)^{\alpha_\nu} \ \left(\frac{E_\nu}{\mathrm{GeV}}\right)^{-\alpha_\nu} 
                & \text{for $E_\nu < \epsilon_{\nu,1}$} \\
             \left(\frac{\epsilon_{\nu,1}}{\mathrm{GeV}}\right)^{\beta_\nu} \ \left(\frac{E_\nu}{\mathrm{GeV}}\right)^{-\beta_\nu} & \text{for $\epsilon_{\nu,1} \le E_\nu <
                \epsilon_{\nu,2}$} \\
             \left(\frac{\epsilon_{\nu,1}}{\mathrm{GeV}}\right)^{\beta_\nu} \ \left(\frac{\epsilon_{\nu,2}}{\mathrm{GeV}}\right)^{\gamma_\nu-\beta_\nu} \ \left(\frac{E_\nu}{\mathrm{GeV}}\right)^{-\gamma_\nu} 
                & \text{for $E_\nu \ge \epsilon_{\nu,2}$}
            \end{cases} 
\label{eq:nuspec}\\[2mm]
\epsilon_1 &= 7\times 10^{5}\,\mathrm{GeV} \ \frac{1}{(1+z)^2}
              \left(\frac{\Gamma_\mathrm{jet}}{10^{2.5}}\right)^2
              \left( \frac{\mathrm{MeV}}{\epsilon_\gamma} \right)\\[2mm]
\epsilon_2 &= 10^7 \, \mathrm{GeV} \ \frac{1}{1+z} \ 
              \sqrt{\frac{\epsilon_e}{\epsilon_B}} \ 
              \left(\frac{\Gamma_\mathrm{jet}}{10^{2.5}}\right)^4
              \left( \frac{t_\mathrm{var}}{0.01\,\mathrm{s}} \right) \ 
              \sqrt{\frac{10^{52}\,\mathrm{erg\,s}^{-1}}{L_\gamma^\mathrm{iso}}}
              \\[2mm]
\alpha_\nu &= 3 - \beta_\gamma \quad , \quad \beta_\nu \ = \ 3 -
\alpha_\gamma \quad , \quad \gamma_\nu = \beta_\nu + 2 
\label{eq:nuIndices}\\[2mm]
\frac{\Delta R}{\lambda_{p\gamma}} &=
	\left( \frac{L_\gamma^\mathrm{iso}}{10^{52}\,\mathrm{erg\,s}^{-1}} \right) \ 
	\left( \frac{0.01\,\mathrm{s}}{t_\mathrm{var}} \right) \ 
	\left(\frac{10^{2.5}}{\Gamma_\mathrm{jet}}\right)^4 \ 
        \left( \frac{\mathrm{MeV}}{\epsilon_\gamma} \right)\\[2mm]
\int_{0}^\infty \mathrm{d}E_\nu &\ E_\nu F_\nu(E_\nu) = 
	\frac{1}{8} \ 
        \frac{1}{f_e} \ 
	\left(1 - (1-\langle x_{p\rightarrow\pi} \rangle)^{\Delta R/\lambda_{p\gamma}}\right)
	\int_{1\,\mathrm{keV}}^{10\,\mathrm{MeV}} \mathrm{d}E_\gamma \ E_\gamma F_\gamma(E_\gamma) 
\label{eq:nuIntegral}
\end{align}\end{linenomath}
\tightlist
\item Parameters of the $\gamma$-ray spectrum $F_\gamma(E_\gamma)$: 
\tightlist
\item[-] $\epsilon_\gamma$: break energy;
\item[-] $\alpha_\gamma$: spectrum index before break energy;
\item[-] $\beta_\gamma$: spectrum index after break energy;
\item[-] ${\cal F}_\gamma$: measured fluence in $\gamma$-rays integrated
over the energy range given in the GCN circulars and reports  \citep{web:gcn:homepage};
\item[-] $f_\gamma$: normalization; obtained from integral of Equation (\ref{eq:gammaIntegral}).
\listend
\item  Parameters of the neutrino spectrum $F_\nu(E_\nu)$:
\tightlist
\item[-] $\epsilon_1$: first break energy;
\item[-] $\epsilon_2$: second break energy;
\item[-] $\alpha_\nu$: spectrum index before first break energy;
\item[-] $\beta_\nu$: spectrum index between frist and second break energy;
\item[-] $\gamma_\nu$: spectrum index after second break energy;
\item[-] $f_\nu$: normalization; obtained from integral of Equation (\ref{eq:nuIntegral}).
\listend
\item $z$: redshift of GRB;
\item $\epsilon_e$: fraction of jet energy in electrons;
\item $\epsilon_B$: fraction of jet energy in magnetic field;
\item $f_e$: ratio between energy in electrons and protons;
\item $L_\gamma^\mathrm{iso}$: isotropic luminosity of the GRB;
\item $t_\mathrm{var}$: variability of the $\gamma$-ray light curve of the GRB;
\item $\Gamma_\mathrm{jet}$: Lorentz boost factor of the jet.
\listend

The expression $1 - (1-\langle x_{p\rightarrow\pi} \rangle)^{\Delta
R/\lambda_{p\gamma}}$ in Equation (\ref{eq:nuIntegral}) estimates the
overall fraction of the proton energy going into pions from the size
of the shock, $\Delta R$, and the mean free path of a proton for
photomeson interactions, $\lambda_{p\gamma}$. Here, $\langle
x_{p\rightarrow\pi} \rangle = 0.2$ is the average fraction of proton
energy transferred to a pion in a single interaction. The expression
ensures that the transferred energy fraction is $\le 1$. The
calculations are insensitive to the beaming effect caused by a narrow
opening angle of the jet as all formulae contain the isotropic
luminosity in conjunction with a $4\pi$ shell geometry,
i.e. effectively use luminosity per steradian. For example, the target
photon density used to calculate $N_\mathrm{int}$ is given by
$n_\gamma \propto L_\gamma^\mathrm{iso} / 4\pi R^2$, where $R$ is the
distance of the shock region from the central black hole.

%% The Bibliography %%

%\bibliographystyle{myapj}
%{\raggedright\bibliography{%
%	myabrv.bib,%
%	ref.bib%
%	}}

\end{document}